\documentclass[12pt]{article}
\usepackage{amsmath,amssymb}
\usepackage{bm,color}
\usepackage[ruled,noend,linesnumbered,figure]{algorithm2e}
\usepackage{graphicx}

\usepackage[english]{babel}
\usepackage{wrapfig}
\usepackage{subfig}
\usepackage{url}
\usepackage{multirow}
\usepackage{tabularx}
\usepackage{hyphenat}
\usepackage{color}

\begin{document}

\title{Dynamic screening of a localized hole during photoemission from a metal cluster}

\author{Natalia E. Koval$^{1,*}$\\
E-mail: natalia$\_$koval$@$ehu.es\\
$^*$ Corresponding author\\
\\
Daniel S\'{a}nchez-Portal$^{1,2}$\\
E-mail: sqbsapod$@$ehu.es\\
\\
Andrey G. Borisov$^3$\\
E-mail: andrei.borissov$@$u-psud.fr\\
\\
Ricardo D\'{i}ez Mui\~{n}o$^{1,2}$\\
E-mail: rdm$@$ehu.es\\
\\
$^1$ Centro de F\'{\i}sica de Materiales
CFM/MPC (CSIC-UPV/EHU), \\ Paseo Manuel de Lardizabal 5,
20018 San Sebasti\'an, Spain \\
$^2$ Donostia International Physics Center DIPC, \\
Paseo Manuel de Lardizabal 4, 20018 San Sebasti\'an,
Spain \\
$^3$ Institut des Sciences Mol\'eculaires d'Orsay, ISMO,
\\ Unit\'e de Recherches CNRS-Universit\'e Paris-Sud UMR 8214, \\
B{\^a}timent 351, Universit\'e Paris-Sud, F-91405 Orsay Cedex, France
}

\maketitle

\begin{abstract}

Recent advances in attosecond spectroscopy techniques have fueled the interest in the
theoretical description of electronic
processes taking place in the subfemtosecond time scale.
Here we study
the coupled dynamic screening of a
localized hole and a photoelectron emitted from
a metal cluster using a semi-classical model.
Electron density dynamics in the cluster is
calculated with Time-Dependent Density Functional Theory and the motion of the
photoemitted electron is described classically.
We show that the dynamic screening of the hole by the cluster electrons affects the
motion of the photoemitted electron.
At the very beginning of its trajectory, the photoemitted electron interacts
with the cluster electrons that pile up to screen the hole. Within our
model, this gives
rise to a significant reduction of the energy lost by the photoelectron.  Thus, 
this is a velocity dependent effect that should be accounted for
when calculating the average losses suffered by photoemitted electrons in metals.

\end{abstract}

\section*{Keywords}

Spherical jellium cluster, TDDFT, dynamic screening, photoemission

\section{Introduction}

\

Photoemission spectroscopy is one of the most important techniques
used to study the structure of molecules, surfaces and
solids~\cite{Schattke}. It is based on the photoelectric effect, discovered
more than 100 years ago by H. Hertz. Later, in 1905, Albert Einstein explained this effect
as a quantum phenomenon, based on the emission of electrons from a target following the
absorption of a photon ("quantum of light"). Photoemission spectroscopy has significantly contributed to the
understanding of fundamental principles in solid state physics.

In recent years, progress in laser technology has made possible the
development of photoemission spectroscopy in the attosecond range (1 as$=10^{-18}$ s)~\cite{Cavalieri}.
Attosecond techniques permit access to the time scale of electron motion in atoms, molecules,
and solids.
Due to this experimental advance, there is a growing interest in the theoretical description
of the dynamical electronic processes taking place
in the subfemtosecond time scale~\cite{Ricardo,Kazansky,Lemell,Krasovskii}.

In the present work, we study the electron dynamics during photoemission
from small metallic clusters (or nanoparticles).
Clusters represent a bridge
between individual atoms and solid state materials \cite{Alonso}.
Due to their large surface to bulk ratio, small metal clusters can exhibit rather
unique features.
For example, they frequently present interesting catalytic properties.
Our choice of a finite size system as target in the photoemission process
simplifies the theoretical analysis, but
some of our conclusions are expected to remain valid in extended systems
such as metal surfaces.

We consider the case where one of the atoms in the metallic cluster
undergoes core-electron photoemission.
We focus our attention on the combined dynamic screening of a static localized
core-hole and the photoemitted electron. We show that the presence of the hole
left behind affects the many-body electronic
dynamics in the cluster and therefore the emission dynamics of the photoelectron.
For the description of the many-body response of the valence electrons
in the cluster we use time-dependent
density functional theory (TDDFT) -- an \textit{ab-initio} quantum-mechanical
method.
Our TDDFT methodology has already been successfully
applied to study the dynamic screening of
charges in finite-size systems~\cite{Borisov2, Borisov3} and to the calculation
of the energy transfer between particles and small
gas-phase clusters~\cite{Quijada1, Quijada2}.
In the present calculations
the motion of the photoemitted electron is described classically. This
approximation is
justified, provided that typical energies of the
photons are in the 100 eV range (XUV), resulting in
relatively high energies of the photoemitted
electrons, as the ones considered here. To analyze the role of the many-body
screening effects we perform calculations using various approximations for
the classical trajectory of the electron, including constant velocity studies
and calculations with and without direct interaction between the ejected
electron and the hole left behind.

The problem we are addressing here has a long history in condensed matter
physics. The dynamic relaxation of the Fermi sea after creation of a hole
was analyzed in the context of X-ray photoemission by several
authors \cite{Gadzuk,Friedel}. Within the framework
of linear response theory, Noguera \emph{et al.} showed that the effective
interaction between the core-hole and
the photoemitted electron changes continuously from a statically
screened potential for low-energy electrons to a completely unscreened potential for high-energy
electrons ~\cite{Friedel}. They also showed that the double screening of hole and electron can
occur with or without creation of plasmons according to the kinetic energy of the
emitted electron.
Here we
go beyond linear theory in
the description of the dynamic screening of charges in the photoemission process
by using propagation of electronic wave-packets with TDDFT
to compute the response of the valence electrons.

\section{Theory}

\

In the present study metallic clusters are described
using a spherical jellium model (JM). Despite its simplicity, the JM
can be very useful in the interpretation of photoemission data
from metal clusters, as recently shown in Ref.~\cite{Jankala}.
In the JM, the core ions are substituted by an
homogeneous background of positive charge with a density defined by

\begin{equation}\label{posit-dens}
n_0^{+}({\bf r})=n_{0}(r_{s})\Theta(R_{cl}-r),
\end{equation}
where $ R_{cl} $ is the radius of the cluster, $ \Theta(x) $ is the Heaviside
step-function and $ n_{0}(r_{s}) $ is the constant bulk
density, which depends only on the Wigner-Seitz radius $ r_{s} $
($1/n_{0}=4 \pi r_{s}^3/3$) ~\cite{Ashcroft}. The latter is
the only parameter in the JM.
The number of electrons in a neutral cluster is $
N =\left( R_{cl}/r_{s} \right)^3$.
For simplicity, we only consider closed-shell clusters in our calculations.

In order to obtain the ground state electronic density of the
cluster $n({\bf r})$ we use the spin-restricted
density functional theory (DFT)~\cite{HK} and solve Kohn-Sham (KS)
equations~\cite{KS}:

\begin{equation}\label{KS-eq}
\left \{-\frac{1}{2} \nabla^2 + V_{eff}([n],{\bf r}) \right \} \varphi_{i} = \varepsilon_{i} \varphi_{i},
\end{equation}
where $ \varepsilon_{i} $ are the eigenvalues of the KS equations and $\varphi_{i}  $ are the one-electron
wave functions. Please, notice that, unless otherwise specified,
we use Hartree atomic units (a.u.) throughout the paper.
The effective potential is composed of three
terms:
\begin{equation}\label{eff_pot}
 V_{eff}([n],{\bf r})=V_{ext}({\bf r})+V_{H}({\bf r})+V_{xc}({\bf r}),
 \end{equation}
where $ V_{ext}({\bf r}) $ is the external potential created by the positive background,
$ V_{H}({\bf r}) $ is the Hartree (or Coulomb) potential created by the electronic density,
and $V_{xc}({\bf r})$
is the exchange-correlation potential, calculated in our case in the local-density approximation (LDA)
with the Perdew-Zunger parametrization of Ceperley-Alder exchange
and correlation potential~\cite{Perdew}.

\begin{figure}[h!]
 \centering
      \includegraphics[width=1.0\textwidth]{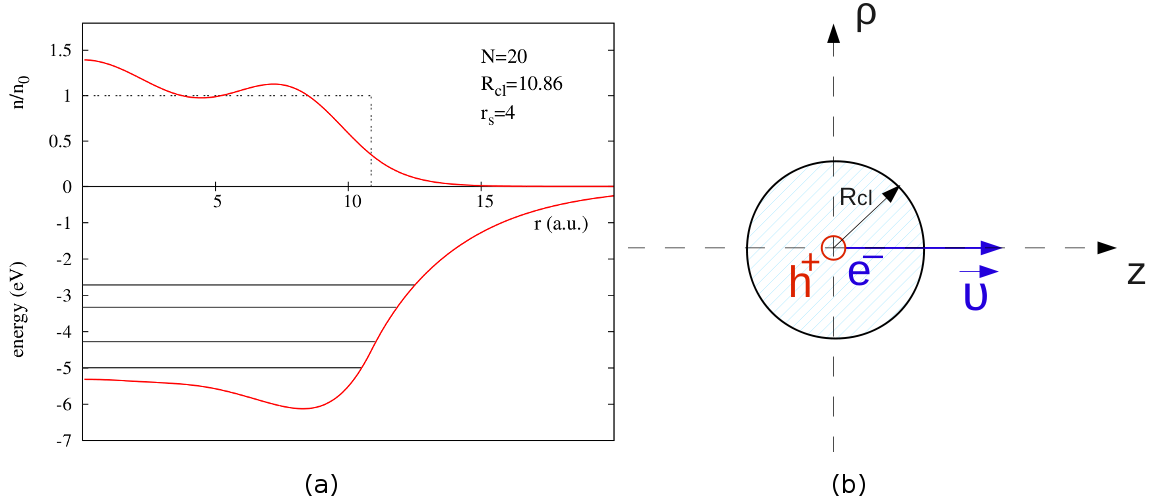}
       \caption{a)
Upper panel, ground state electronic density
in units of the positive background density (dashed line, $r_s$=4) for a cluster
containing 20 electrons.
Lower panel,  the corresponding effective Kohn-Sham potential
and occupied energy levels. b) Sketch of the photoemission process. An electron and a hole
       are created at the center of the spherical cluster
at $t=0$. Both are represented by classical point particle
and the electron
       starts to move along the $z$-axis
       with velocity $\upsilon_0$. }
       \label{scheme}
\end{figure}

The electronic density $n({\bf r})$ is given by a sum over occupied wave functions:
\begin{equation}\label{dens}
n({\bf r})= 2 \sum_{i \in occ} |\varphi_{i}({\bf r})|^2,
\end{equation}
where the factor 2 stands for spin degeneracy.
Results for the ground state of small metal clusters have been thoroughly discussed in the
literature ~\cite{Ekardt,Heer,Knight}.
The effective KS potential and the
ground state electronic density
for one of the clusters considered here
are shown in Figure~\ref{scheme}(a).

For the description of the photoemission process we use a semi-classical model.
We consider fast photoemitted electrons, moving with velocities much
higher than the Fermi velocity of the cluster electrons. Therefore, the movement
of the photoelectron can be represented classically. In parallel, valence electron
dynamics in the cluster is investigated by means of TDDFT~\cite{RG}. In TDDFT, the
evolution of the electronic density $n({\bf r})$ in response to the field of the
moving electron is calculated solving the time-dependent KS equations:
\begin{equation}\label{TDKSeq}
\mathrm{i} \frac{\partial}{\partial t} \varphi_{i} ({\bf r},t)=
\left \{-\frac{1}{2} \nabla^2 + V_{ext}({\bf r})+V_{H}({\bf r},t)+
V_{xc}({\bf r},t) + \triangle V({\bf r},t)  \right \} \varphi_{i}({\bf r},t),
\end{equation}
where $\triangle V({\bf r},t)$ is the change of the
external
potential due to the
photoemission process (see below).
The exchange-correlation potential $ V_{xc} $ is calculated with the standard
adiabatic local
density approximation (ALDA) with the parametrization of Ref.~\cite{Perdew}.

The time-evolving electronic density of the
excited cluster is
obtained from the time-dependent
KS orbitals $ \varphi_{i}({\bf r},t) $, in a way similar to Eq. \eqref{dens}.
The time-dependent KS wave functions are obtained by propagating the initial wave
functions $ \varphi_{i}({\bf r},t_0)  = \varphi_{i}({\bf r})$ using the
split-operator technique.
Due to the presence of the photoemitted electron the problem loses its spherical
symmetry and the use of cylindrical coordinates ($\rho, z$) becomes
necessary. A detailed description of the
numerical procedure can be found in
Refs.~\cite{Borisov1,Borisov-WPP,Chulkov}.

In our model, we do not consider explicitly the interaction with
an external electromagnetic field.
Thus, an electron with high kinetic energy and a static hole are created at $t=0$
at the center of the spherical cluster.
The scheme of the
process is shown in Figure~\ref{scheme}(b).
The photoemitted electron (\textit{el}) is modeled as a negative
point charge that
moves along the $z$-axis
$[\rho=0,~z_{el}(t)]$.
It is worth noting that this
photoemitted electron is not one of the cluster valence electrons,
but an extra electron coming from the inner shell of a hypothetical
atom sited in the center of the cluster.
The electron motion is calculated in two different
approximations as described in detail below.
The hole (\textit{h}) is
represented by
a positive
point charge at a fixed position $[\rho=0,~z_{h}=0]$.
The potential created by these charges and acting on the rest of the
electrons in the cluster is $\triangle V = V_{el}+V_{h}$, where
\begin{equation}\label{pot-ext-el}
V_{el}=\frac{1}{[(z_{el}(t)-z)^2+\rho^2]^{1/2}} \Theta(t),
\end{equation}
and
\begin{equation}\label{pot-ext-hole}
V_{h}=-\frac{1}{[z^2+\rho^2]^{1/2}} \Theta(t).
\end{equation}

To address the effect of the many-body dynamics in
the cluster on the energy loss experienced by the ejected electron in
well-defined conditions, we first study a simplified case
in which the photoemitted electron is assumed to move
with constant velocity $\upsilon$, i.e.,
$z_{el}(t)=\upsilon t$.
This allows us to isolate the effects related to the
dynamics of the screening processes from other possible effects associated with
the details of the trajectory. Here, $\upsilon$ corresponds to the final
velocity of the electron if the photoemission process would
take place in vacuum, which is considered as a good approximation
for the average electron velocity during its movement through
the cluster. Thus, the direct interaction
between the electron and the hole is not taken into account in
this case. However, as we will see below, the screening of the hole
still has an important influence on the energy loss by the photoemitted electron.
In this case the energy loss is calculated from the integral
\begin{equation}
E_{loss}= -\upsilon\int^{\infty}_{0} F^{cls}_z(t)~dt ,
\label{eloss}
\end{equation}
where
\begin{equation}\label{force}
F^{cls}_{z}(t) = 2\pi \int d\rho dz~\rho
\frac{n(\rho,z,t)-n^{+}_{0}(\rho,z)}{[(z_{el}(t)-z)^2+\rho^2]^{3/2}}
[z_{el}(t)-z]
\end{equation}
is the $z$-component of the force created by
the cluster on the emitted electron.
It is important
to note that $E_{loss}$ includes the energy necessary to eject the
electron from the cluster (an adiabatic contribution), as well as
non-adiabatic contributions
due to the creation of electronic excitations in the cluster during the
emission process.

In a second step, we perform a more refined treatment in which
the direct electron-hole interaction is included and
the trajectory $z_{el}(t)$ is calculated using
the classical equations of motion:
\begin{equation}\label{coord}
d z_{el}/dt= \upsilon(t),~~~~z_{el}(t=0)=0
\end{equation}
\begin{equation}\label{vel}
d\upsilon/dt= F^{tot}_{z}(t),~~~~\upsilon(t=0)= \upsilon_{0}.
\end{equation}
In Eqs.(\ref{coord}) and (\ref{vel}), $F^{tot}_{z}$ is the
total force
felt by the moving electron:
\begin{equation}\label{totalforce}
F^{tot}_{z}(t) = F^{cls}_z(t)
-\frac{z_{el}(t)}{[z_{el}(t)^{2}+\alpha^2]^{3/2}}.
\end{equation}
The first term corresponds to the interaction with the cluster
given by Eq.(\ref{force}).
The
second term stands for the force due to the interaction between
photoemitted electron and the core hole left behind.
The electron-hole interaction in our study is given by the regularized
Coulomb potential $V_{el-h}$:
\begin{equation}\label{e-h-int}
V_{el-h}=-\dfrac{1}{\sqrt{z_{el}(t)^2+\alpha^2}},
\end{equation}
We use $\alpha^2=0.5$
to avoid divergence at time $ t=0 $.

\

\section{Results and discussion}

\

In what follows, we center our discussion on the force experienced by the
photoemitted electron due to the interaction with the cluster.
This force is given by Eq.\eqref{force} and allows
to study important aspects of the electron density dynamics in the
cluster.  We study
such force
in the two approximations mentioned in the previous section for the motion
of the (classical) photoemitted electron. In the first approximation, in which
the electron moves with constant velocity, we can conveniently identify the effect
of the hole screening on the movement of the photoemitted electron.
To quantify the effect of the dynamic screening of the hole
we calculate the work
performed by
the force $F^{cls}_{z}$
along the electron
trajectory. This quantity is directly linked with
the energy loss of the ejected
particle.
In the second approximation, the velocity  of the electron is allowed
to vary according to Newton laws. This approximation
might be closer to the real
photoemission process.
Also in this case we find an important influence of the
hole-screening dynamics on the force experienced by the emitted
electron and, thus, on the energy loss during the photoemission process.

\

\subsection{Constant velocity approximation}

\

In the constant velocity approximation,
we calculate the cluster induced force in two different cases,
namely with a localized hole at the center of the cluster
(potentials in Eqs. \eqref{pot-ext-el} and
\eqref{pot-ext-hole} are included in the calculations)
and without the hole
(only the potential in Eq. \eqref{pot-ext-el} is included).
In this approximation the direct interaction between hole and electron is not included.
In spite of this, we find that
the presence of the hole modifies the electron dynamics in the cluster because
there are two different charges to be screened. We have studied four Na clusters
($r_{s}=4$) with 20, 58, 106 and 556 electrons.
We consider three different velocities of the photoemitted electron: 1 a.u, 1.5 a.u. and 2.5 a.u.
\begin{figure}[h!]
  \centering
      \includegraphics[width=1.0\textwidth]{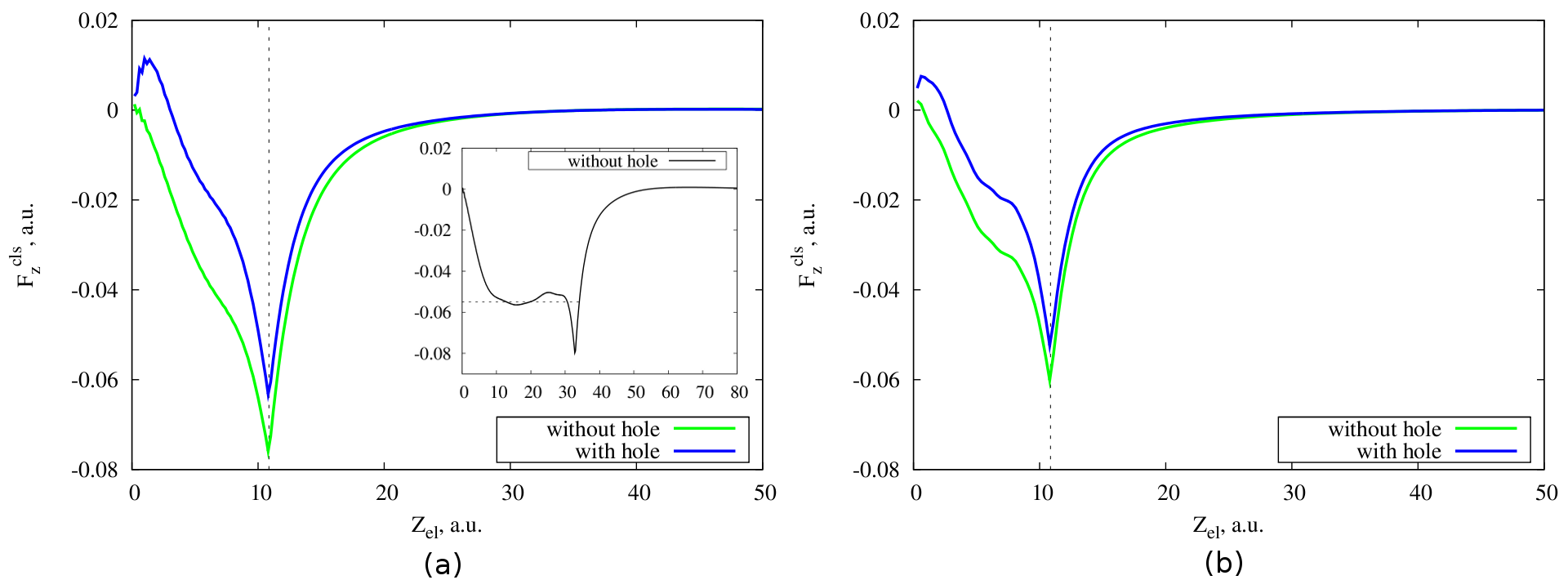}
       \caption{Cluster induced force [Eq.~\eqref{force}] acting on the electron moving away from
       the center of the cluster (N=20, $r_{s}$=4) as a function of the electron position.
 We consider the cases without and with a localized hole at the center of the cluster.
The electron moves
with a constant velocity: a) $ \upsilon = 1$ a.u.
(Inset: Cluster induced force
for a larger cluster with 556 electrons, the hole is not included.
The horizontal dotted line corresponds
to the stopping power in an homogeneous electron gas with $r_s$=4.)
 and b) $ \upsilon = 1.5$ a.u.
Vertical dashed lines correspond to the radius of the cluster, $R_{cl}=10.86$ a.u. }
       \label{force-1st}
\end{figure}
Figure~\ref{force-1st} shows the cluster induced force acting on the
photoemitted electron as a function
of the
electron position in the cases $ \upsilon = 1$ a.u. and $ \upsilon = 1.5$ a.u.
for the small cluster with 20 electrons.
From the present results it follows that, in the case
of the presence of the hole, the cluster induced force on the photoemitted
electron has positive value at short times for the two chosen velocities. This indicates that
the cluster response tends to accelerate the electron
at the very beginning of its movement.
The repulsive cluster induced force is related to the screening of the hole.
More precisely, at the beginning of its movement,
when close to the hole, the
electron is repelled from the hole vicinity by the cluster
electrons that arrive to screen the hole.
When the hole is not included in the calculation, the above effect
is not observed
and the emitted electron is
mainly decelerated all along its trajectory~\cite{post-force}.
This deceleration is due to two effects. First, within the cluster, the electron
suffers the stopping characteristic of any charged
particle moving in an electron gas~\cite{Quijada2}. Second, as
the electron approaches the cluster surface, we can clearly see the contribution
to the forces associated with overcoming the surface potential step.

These two decelerating contributions are difficult to disentangle for
very small clusters, like those in the main panels of Figure~\ref{force-1st}.
However, the force experienced by a particle moving inside
a large jellium cluster reaches a  stationary regime and oscillates around a mean value.
This can be seen in the inset of Figure~\ref{force-1st}~(a) for a cluster containing
556 electrons.
The mean value of the force is the so-called
stopping power and only depends on the electron
density. In our case, for $r_s$=4, is around 0.055 a.u.
for a negatively charged particle moving with a velocity of 1 a.u.~\cite{Quijada1}.

\begin{figure}[h!]
\centering
 \includegraphics[width=1.0\textwidth]{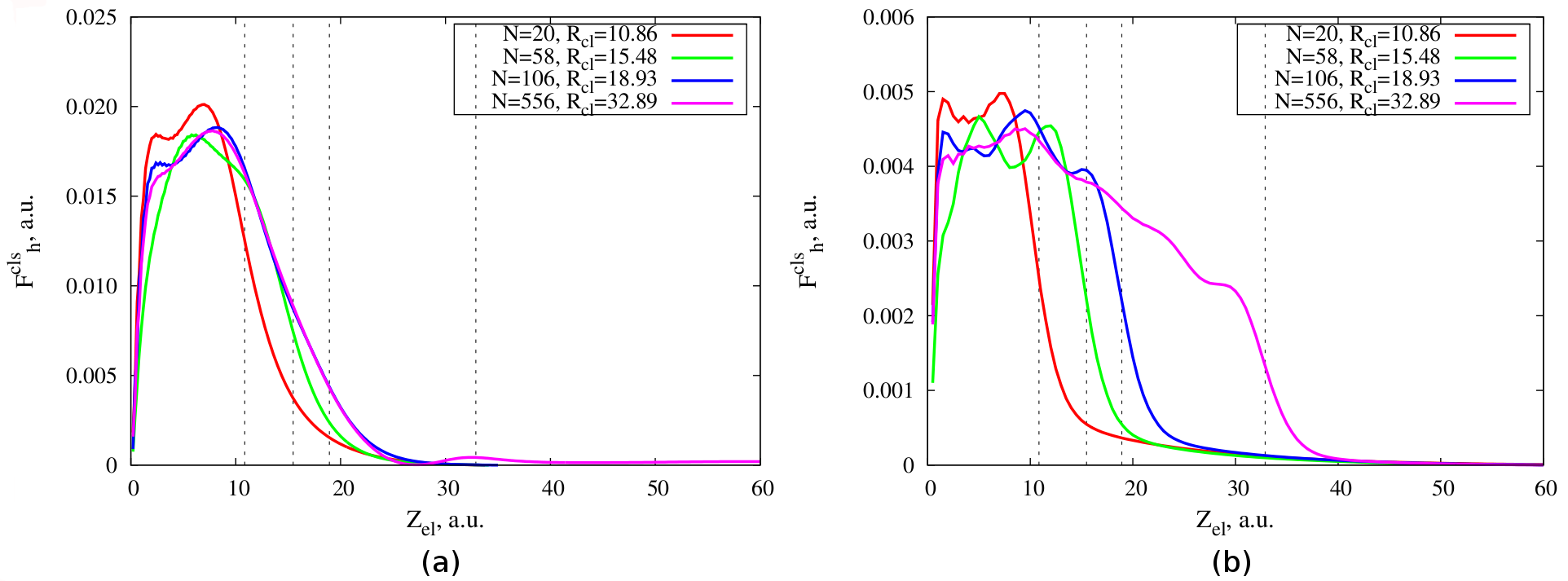}
  \caption{
Component of the cluster induced force acting on the moving electron
       due to the dynamic screening of the hole by the electronic charge
      of the cluster, $F^{cls}_{h}(z)$ .
       Results are shown as a function of the electron position for
       jellium clusters ($r_{s}$=4) of different size comprising
       $N=20,\ 58,\ 106$, and $556$ electrons. Electron velocities are: a) $ \upsilon = 1$ a.u,
       b) $ \upsilon = 2.5$ a.u. Vertical dashed lines correspond to the radii of the clusters. }
\label{fig:dforce}
\end{figure}

The influence of the hole screening on the moving electron can be
better analyzed if we look at the difference of forces shown in the
Figure~\ref{force-1st}: $F^{cls}_{h}(z)=F^{cls}_{h,el}(z) - F^{cls}_{el}(z)$.
Here, $F^{cls}_{h,el}(z)$ ($F^{cls}_{el}$) is the cluster induced force
on the photoemitted electron calculated with (without) explicit inclusion of
the positive point charge at the center of the cluster.
With this definition, $F^{cls}_{h}(z)$ is the force felt by the photoemitted
electron due specifically to the cloud of electronic density that
dynamically screens the hole. This quantity is shown in Figure~\ref{fig:dforce}
for all clusters considered and for two different velocities.

As it is seen from the graphs, the effect of the screening of the hole is larger in the case of
the
smaller electron velocity 1 a.u. - Figure~\ref{fig:dforce}(a).
This is related to the time that the photoelectron spends in the neighborhood of the hole and to the
characteristic time of the hole screening. The slow photoelectron stays near the hole long enough
for the screening of the hole to be performed.
Therefore, it experiences a large force due to the piling up of
electronic charge around the hole.
The fast electron, however, leaves the hole at short times, which are not
enough for a significant
piling up
of screening charge.
Hence,
the effect associated with the hole screening
becomes smaller the higher the electron velocity.
It is worth noting that,
for the slow electron, $F^{cls}_{h}(z)$
is almost identical for the
two largest clusters considered here and it is very small for
$z_{el} > 25$~a.u. Both observations indicate that the screening of the hole is well
established and basically reaches its stationary value at the corresponding time scale.
For the faster electron, however, the value of $F^{cls}_{h}(z)$ at large $z_{el}$
is different for different cluster radii.
This is linked to the time-evolution of the screening
density which still goes on by the time
the electron reaches the cluster boundary.
These conclusions are corroborated by the induced electron density dynamics plots
(Figure~\ref{fig:dens-z-t-cont}) discussed below.

In order to quantify the effect of the hole screening, we calculate the
electron energy loss due to the interaction with the cluster electrons for
the case of $N=20$ and for the velocities 1 a.u, 1.5 a.u., and 2.5 a.u. of the photoemitted
electron.
The energy loss $E_{loss}$, given by Eq.~\eqref{eloss}, is defined
as the work performed by
the cluster induced force acting on the moving electron.
The results are summarized in Table~\ref{loss}, in which the difference in
the cluster induced energy loss
$\Delta E_{loss}$,
with and without the hole, is also shown.

\begin{table}[h!]
\begin{center}
\caption{
Estimated energy loss [Eq.~\eqref{eloss}] for a photoemitted
electron as a function of its velocity. The electron is assumed to follow
a trajectory with constant velocity.
The difference in energy loss $\Delta E_{loss}$
for the case of absence and presence of a hole
in the center of the cluster is also given. } 
\begin{tabular}{cc|c|c|c|l}
\cline{1-5}
\multicolumn{2}{|c|}{N=20} & $\upsilon=1\ a.u.$ & $\upsilon=1.5\ a.u.$ & $\upsilon=2.5\ a.u.$  \\ \hline \cline{1-5}
\multicolumn{1}{|c|}{\multirow{2}{*}{$ E_{loss}, a.u.$}} &
\multicolumn{1}{|c|}{without hole} & 0.60 & 0.44 & 0.22  \\ \cline{2-5}
\multicolumn{1}{|c|}{}                                                          &
\multicolumn{1}{|c|}{with hole} & 0.37 & 0.30 & 0.16  \\ \hline \cline{1-5}
\multicolumn{2}{|c|}{$\Delta E_{loss}$ , a.u.} & 0.23 & 0.14 & 0.06  \\ \hline \cline{1-5}
\end{tabular}
\label{loss}
\end{center}
\end{table}
The presence of the hole
reduces the cluster induced energy loss for all velocities.
The value of $\Delta E_{loss}$ also shows
that the effect of the hole screening
is more significant the slower the electron.
The energy loss of the electron moving at 1 a.u.
decreases almost
by a factor of 2 when we include the hole screening in the process.
An interesting consequence is that, at low velocities, the
effects associated
with the hole screening might become crucial to determine
if the photoemission process can indeed take place or not. For example, the kinetic energy
of the slowest electron considered in Table~\ref{loss} is 0.5 a.u.. Since
the energy loss  in the case without hole is 0.6 a.u.,
this electron cannot be photoemitted from
the cluster. However, in the presence of the hole the photoemission
becomes possible.

The study of the electron dynamics during photoemission in the constant velocity
approximation leads us to two conclusions: 1) the screening of the hole by the
cluster electrons leads to a repulsive
(accelerating)
force acting on the photoemitted electron at the beginning of its movement;
2) the effect of the hole screening is reduced
for faster (more energetic) photoemitted electrons.

\

\subsection{Varying velocity approximation}

\

The results discussed so far are obtained using a simple model in which the
photoemitted electron moves with a constant velocity. In a real photoemission
process, however, the velocity varies due to the different
elastic and inelastic forces acting on the electron. In order to be sure that
none of the effects discussed above is an artifact of the model and to prove our
conclusions, we simulate the photoemission process in a more realistic second
approximation. In this approximation, the velocity and coordinate of the
electron are dependent on time, according to
equations \eqref{coord}, \eqref{vel}, and \eqref{totalforce}. Electron and hole
interact via a regularized Coulomb potential \eqref{e-h-int}.
\begin{figure}[h!]
  \centering
      \includegraphics[width=0.6\textwidth]{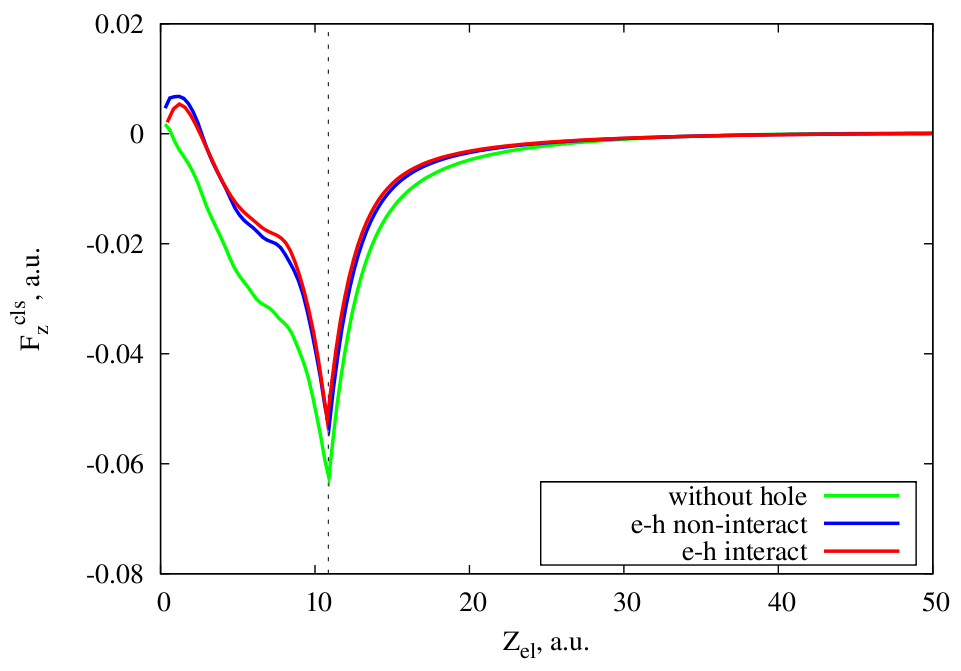}
       \caption{Cluster induced force
       [Eq.~\eqref{force}]
       acting on the electron moving
       away from the center of the cluster (N=20, $r_{s}=4$), as a function
       of the electronic position in three cases: 1) There is no hole
       at the center of the cluster and the initial
       velocity of the electron is 1.5 a.u. (green line); 2) There is a
       hole at the center of the cluster, but the photoemitted electron does not
       interact directly with the hole. The initial velocity of the electron
       is 1.5 a.u. (red line); 3) There is a
       hole at the center
       of the cluster and
       a direct electron-hole interaction (Eq.~\ref{e-h-int}) is included.
       The initial velocity of the
       electron is 2.25 a.u. (red line). The vertical dashed line shows the radius of the cluster,
       $R_{cl}=10.86$ a.u.}
       \label{force-2nd}
\end{figure}

In Figure~\ref{force-2nd} we show the results for the small cluster with 20 electrons
in this varying velocity approximation.
The force felt by the moving electron due to the interaction with the cluster electrons
[Eq.~\eqref{force}]
is calculated for three
different cases. In the first case, we do not include the hole in the cluster.
In the second case, there is a hole to be screened, but we do not include the direct
interaction between the hole and the photoelectron
when performing the
trajectory calculation.
In the third case, the direct interaction between photoelectron and hole
is included.
In the first two situations, the initial velocity of the electron is
set to 1.5 a.u, while in the third situation is
set to 2.25 a.u.
This difference in velocity corresponds to the energy of the electron-hole
interaction. Indeed, with the parameter $\alpha^2=0.5$ a.u.
used here in Eq.\eqref{e-h-int}, the ``binding energy"~\cite{foot}
of the electron is around 1.4 a.u.
Taking into account this binding energy, the initial velocity 2.25 a.u.
for an electron photoemitted in vacuum and interacting with the hole leads to a final
velocity $\sim$~1.5~a.u. Therefore, our
choice of initial velocities allows a direct comparison of
the two cases, with and without direct
electron-hole interaction in the case of photoemission inside the cluster.

One can see from Figure~\ref{force-2nd} that the behavior
of the cluster induced force in this more realistic model is similar to that
of the simple model considered before (Figure~\ref{force-1st}). Whenever the hole
screening is taken into account, there is an acceleration force
acting
on the electron
at the beginning of its trip. Similarly to the constant velocity approximation,
performing calculations along
a more realistic trajectory with different launch
velocities, we also found that the effect of the hole screening decreases
when the initial velocity of the electron increases.

Moreover, we found that the cluster induced force is very similar
independently on whether electron and hole directly interact with each other or not.
However, this is valid only if the final velocity of the photoemitted
electron interacting with the hole is equal to the final velocity of the electron not
interacting with the hole. This shows that, as far as the final energy of the photoemitted
electron is the same, the cluster induced force acting on the
photoemitted electron is
mainly affected by the presence or absence of the hole screening and not by the details
of the electron trajectory nearby the hole.
\

\subsection{Time evolution of electronic density}

\

\begin{figure}[h!]
  \centering
  \includegraphics[width=1.0\textwidth]{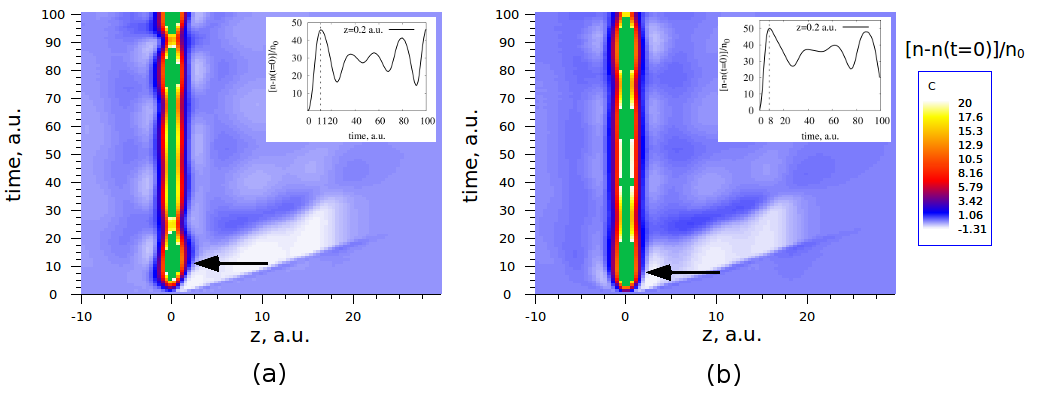}
  \caption[Time evolution of the electronic density of the cluster, non-linearity]
{Time evolution of the electronic density of the spherical cluster in which the hole and the electron are created at time $ t=0 $ at the center of the cluster and the electron moves away along the
positive part of the $z$-axis.
The induced electronic density is shown close
to the symmetry $z$-axis ($\rho$=0.02 a.u.,$z$).
The time evolves along the vertical axis. Color map shows the change in density in units of the background density $n_{0}$. The color scale is
limited to a maximum value of $20$ in order to reveal
  the effects in the regions where the induced density is small. The induced density above this value is shown in green. The actual maximum value of the induced density is about $46-50$ in units of background density. It corresponds to the small $z$ region around
the position of the hole. Panel (a) shows the results of the TDDFT calculation
of the complete system. In panel (b)
the induced density is calculated as a sum of two contributions (see the text for the explanation). Cluster parameters are $r_{s}=4$, $N=106$, $ R_{cl}=18.93$ a.u.
The velocity of the electron is constant and is equal to $ \upsilon=1$ a.u.
Insets: Profile of the plot along the time axis at ($\rho$=0.02 a.u., $z$=0.2 a.u.).
}
\label{fig:dens-z-t-cont}
\end{figure}

Continuing the discussion on the electron density dynamics
in the cluster, we illustrate the effect of the coupling of both processes -- dynamic
screening of the hole and dynamic screening of the moving electron.
Figure~\ref{fig:dens-z-t-cont} shows the time evolution of the electronic density
of the spherical cluster with $N=106$ electrons. The hole and the electron are created
at time $t=0$ at the center of the cluster ($z$=0) and the electron is moving along the
positive part of the $z$-axis with a constant velocity $\upsilon=1$ a.u.
The induced electronic density
close to the $z$-axis $ \Delta n=n(\rho_0,z,t)-n(\rho_0,z,0) $,
where $ \rho_0=0.02 $ a.u, is plotted in units of the background density $n_0$.

Panel (a) shows the results of a direct TDDFT calculation of the
induced electronic density for the cluster with a static hole
at the center and a photoelectron moving away from the center of
the cluster along the $z$-axis.
Panel (b) shows the
induced electronic density obtained as a sum of two different
contributions:

\begin{equation}\label{delta-dens}
\Delta n(t) = \delta n_{h}(t) + \delta n_{el}(t).
\end{equation}
Here, $ \delta n_{h}(t) $ is the TDDFT result for the induced electron
density due to the appearance of only a localized hole at the center of
the cluster. Similarly, $ \delta n_{el}(t) $ is the TDDFT result for the induced
electron density in response to a photoelectron moving from
its center, where no hole is present. Therefore, panel (b) shows a linear
superposition of the electronic charges screening the
static hole and the moving photoelectron.
In the inset of both graphs we show the time evolution of the
electronic density at a given point ($\rho$=0.02 a.u., $z$=0.2 a.u.).

The white area in the main plots shows a depletion of the electronic density in
the cluster that roughly follows the trajectory of the electron. It is due
to the Coulomb repulsion between the moving electron
and the rest of the electrons in the cluster.
Black arrows indicate the time at which the screening of the hole is fully developed, i.e.,
the induced electron density in the close vicinity of the hole roughly
integrates to one. This time is also shown
in the inset of each plot and is equal to 11 a.u.
and 8 a.u. for the cases (a) and (b), respectively.
Thus, there is a delay in the TDDFT screening of the hole
as compared to the linear superposition case.
Moreover,
comparing the charge distribution for negative and positive values of $z$, we
can see a clear asymmetry in the screening charge for the
TDDFT calculation with both hole and electron  simultaneously included. This asymmetry is
absent in panel (b), corresponding to the linear superposition of electron and hole
separate screenings, and clearly indicates that the dynamics of the hole
screening is
affected by the presence and movement of the emitted electron.
Therefore, we can conclude that the TDDFT calculation, considering both the hole
at the center and the electron photoemitted from the center of the cluster, includes a
combined effect of the dynamic screening of both particles in the relaxation processes in
the cluster. This combined effect is also visible in the oscillations of the electronic
density, where the periods of these oscillations are slightly different
for the two cases considered.

\section{Conclusions}

\

In this study a semi-classical model was used to describe the dynamic screening
of a moving photoelectron and that of the localized core-hole left
behind as a result of interaction of XUV pulses with small metal clusters.
The motion of the photoemitted electron is described classically and the electron
dynamics in the clusters is studied using the Time-Dependent Density
Functional Theory, TDDFT.

We have shown that, when the hole is explicitly included in the calculation,
the photoemitted electron is accelerated by the cluster electrons that pile up nearby the
cluster center to dynamically screen the hole.
This effect is observed by comparing the forces acting on the
photoemitted electron due to the interaction with the cluster in which a hole is
present or absent at the center.
In order to quantify the effect of the 
hole screening we have calculated the energy loss
of the photoelectron. We have shown that the presence of the hole
reduces significantly the cluster induced energy loss 
and that this effect is velocity dependent.
The higher is the energy of the photoemitted electron
the smaller is the effect induced by the hole screening.

These conclusions were obtained
using a relatively simple approximation in which the photoemitted electron moves with constant
velocity. The conclusions are proven to remain valid when
the interaction between photoemitted electron and core-hole left behind is included in
the calculation and the velocity of the electron is allowed to vary with
time.
We have illustrated the time evolution of the electron density in the cluster during the
photoemission process and we have shown that the TDDFT calculation allows us to see the coupled
effect of the screening of both the hole and the electron in the relaxation processes inside
the cluster.

The semi-classical model used here allows for a detailed analysis of the effect of the dynamic screening of the hole. However, the simplicity of the model and the classical treatment of the photoemitted electron also prevents a direct translation of our findings to the experimental 
situation. 
Thus, a clear understanding of the implications of the present results for photoemission experiments is still an open question that requires further work.

\section*{Abbreviations}

TDDFT, Time-Dependent Density Functional Theory; XUV, Extreme ultraviolet; JM, Jellium Model; KS, Kohn-Sham.

\section*{Competing interests}

The authors declare that they have no competing interests.

\section*{Authors' contributions}

All authors have made substantial contributions to conception, acquisition and interpretation of data.
All authors have been involved in drafting the manuscript. DSP, AGB and RDM have been revising the manuscript
for important intellectual content and have given final approval of the version to be published.

\section*{Acknowledgements}

NEK acknowledges support from the CSIC JAE-predoc program,
co-financed by the European Science Foundation.
We also acknowledge the support of the Basque Departamento de
Educaci\'on and the UPV/EHU (Grant No. IT-366-07), the
Spanish Ministerio de Econom\'{\i}a y Competitividad
(Grant No. FIS2010-19609-CO2-02) and the
ETORTEK program funded by the Basque
Departamento de Industria and the Diputaci\'on Foral de Gipuzkoa.


\begin{thebibliography}{99}
\bibitem{post-force}
In Figure~\ref{force-1st}(b) one can see that, even in
the calculation without hole, the force
acting on the photoemitted electron
exhibits very small positive values at the beginning
of the electron trajectory. In this case,
this is not related to the
dynamic processes inside the cluster, but to the slightly
inhomogeneous distribution
of the electron density in the small clusters. The ground state electronic density
of a cluster with 20 electrons and $r_s$=4 is shown in
Figure~\ref{scheme}~(a). The electron density shows a maximum at the center, where it is
larger
than the value of the positive background density.
Since we start the dynamical calculations using the ground state density,
this leads to a positive (i.e., repulsive) value of the
initial force acting on a negatively charged particle located
nearby the center of the cluster.
\bibitem{foot} Binding energy here means
the interaction energy when both the electron and the hole
are located at the center of the cluster.
\end{thebibliography}
\end{document}